\documentclass[prl,twocolumn,showpacs,superscriptaddress]{revtex4}
\usepackage{graphicx}
\usepackage{mathrsfs}
\usepackage{amsmath}
\usepackage{amssymb}
\usepackage{amsfonts}
\usepackage[normalem]{ulem}
\usepackage{color}

\begin{document}

\title{Quantum phases of quadrupolar Fermi gases in optical lattices}
%\author{TBD}

\author{S.~G.~Bhongale} \affiliation{School of Physics, Astronomy, and
  Computational Sciences, George Mason University, Fairfax, VA 22030}
%\affiliation{Joint Quantum Institute, National Institute of Standards
%  and Technology \& University of Maryland, Gaithersburg, MD 20899 }
 
\author{Ludwig~Mathey} \affiliation{Zentrum f{\"u}r Optische
  Quantentechnologien and Institut f{\"u}r Laserphysik,
  Universit{\"a}t Hamburg, 22761 Hamburg, Germany}
 
\author{Erhai Zhao} \affiliation{School of Physics, Astronomy, and
  Computational Sciences, George Mason University, Fairfax, VA 22030}

\author{Susanne F. Yelin} \affiliation{Department of Physics, University
  of Connecticut, Storrs, Connecticut 06269} \affiliation{ITAMP,
  Harvard-Smithsonian Center for Astrophysics, 60 Garden Street,
  Cambridge, MA 02138} \affiliation{Department of Physics, Harvard University, 17 Oxford Street,
  Cambridge, MA 02138}
 
\author{Mikhail Lemeshko} \email{mlemeshko@cfa.harvard.edu}
\affiliation{ITAMP, Harvard-Smithsonian Center for Astrophysics, 60
  Garden Street, Cambridge, MA 02138}%
\affiliation{Department of Physics, Harvard University, 17 Oxford Street,
  Cambridge, MA 02138} %

\date{\today}

%\maketitle
\begin{abstract} 
  We introduce a new platform for quantum simulation of many-body
  systems based on nonspherical atoms or molecules with zero dipole moment but  possessing a
  significant value of electric quadrupole moment. We consider a
  quadrupolar Fermi gas trapped in a 2D square optical lattice, and show that
  the peculiar symmetry  and broad tunability of the
  quadrupole-quadrupole interaction results in a rich phase diagram
  encompassing unconventional BCS and charge density wave phases, and
  opens up a perspective to create topological superfluid.  Quadrupolar species, such as metastable
  alkaline-earth atoms and homonuclear molecules, are stable against
  chemical reactions and collapse and are readily available in
  experiment at high densities.
\end{abstract}
\pacs{67.85.-d, 75.30.Fv, 71.10.Fd}

\maketitle
Quantum gases of ultracold atoms have provided a fresh perspective on strongly-correlated many-body states, by establishing a highly tunable environment in which both open questions of solid state physics and novel, 
 previously unobserved, many-body states can be studied~\cite{BlochRMP08}. An important landmark was reached by cooling and trapping dipolar atoms and
molecules, bosonic and fermionic~\cite{NiScience08,
  ChotiaPRL12, DeiglmayrPRL08, GriesmaierPRL05, LuPRL12, AikawaPRL12}, near or into quantum degeneracy, 
which extended the range of features available to quantum
simulation in ultracold atom systems beyond contact interactions. 
 Numerous exotic states such as 
 supersolids, quantum liquid crystals and bond-order solids have been predicted,  
  extended Hubbard models with 3-body interactions, and highly
tunable lattice spin models for quantum
magnetism have been proposed~\cite{BaranovPhysRep08, LahayePfauRPP2009, BarbaraPRL10,
  FregosoPRL09, BhongalePRL12, gorshkov11c}.  
   The crucial feature of the interactions  in dipolar gases
  is their anisotropic and  long-range character tunable with static and radiative fields~\cite{LemeshkoPRA11Optical, LemeshkoPRL12,
  gorshkov11c}, which  
  is key to the intriguing many-body effects that have been predicted.

In this Letter we propose to study quadrupolar quantum gases. This constitutes a
 new class of systems in  ultracold  physics, which can be used
 as a platform for quantum simulation. 
 Quadrupole interactions are most visible for non-polar particles  which 
 possess a significant electric quadrupole moment.  
The angular dependence of the resulting quadrupole-quadrupole
interaction is substantially different compared to the
dipole-dipole one, due to the higher-order  symmetry. 
 For atoms or molecules in an optical lattice, 
 this allows for broad tunability of the
 nearest and next-nearest neighbor
couplings.
 As a concrete example, we discuss metastable
alkaline-earth atoms and homonuclear molecules which have comparatively large quadrupolar moments.
In order to demonstrate rich many-body effects that arise in ensembles of such particles, we derive the quantum phase diagram of a quadrupolar fermionic gas in an optical lattice at half-filling.  We find that several unconventional phases emerge, such as bond order solids  and $p$-wave pairing, and discover the intriguing
possibility of creating topological ground states of
$p_x+ip_y$ symmetry.
  While dipolar quantum gases were also shown to
host novel many-body phases, quadrupolar particles are available in
experiment at higher densities and are stable against chemical
reactions~\cite{OspelkausSci10} and
collapse~\cite{KochNatPhys08}.

 \begin{figure}[b]
\includegraphics[width=\linewidth]{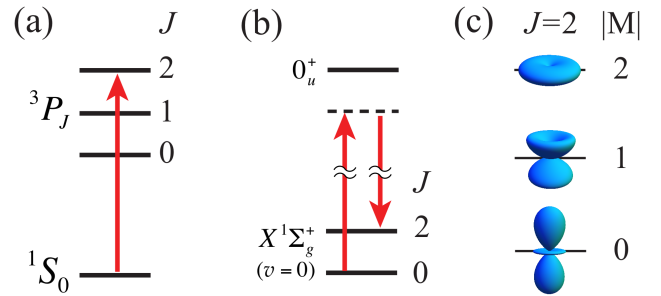} %{levels.pdf}
\caption{
% added "recipe for"
Recipe for 
realization of quadrupolar particles: (a)~with alkaline-earth
  atoms in long-living $^3P_2$ levels; and (b)~with homonuclear
  molecules in rotational states with $J>0$. (c)~Angular ``shape'' of
  quadrupolar particles exemplified by $\vert 2, M \rangle$ states.}
\label{levels}
\end{figure}

In order to determine the quadrupole-quadrupole interaction energy, we consider the potential of a classical quadrupole with moment $q = \int \rho (\vec{r}) r^2 (3 \cos^2\theta-1)d\vec{r}$ located at $\vec{r}_{0}=0$ aligned in $\hat{k}$-direction. Here $\rho(\vec{r})$ is the electron charge density and $\cos\theta \equiv \hat{k} \cdot\hat{r}$ \footnote{In contrast to a dipole which is
analogous to a single-headed arrow ($\uparrow$) pointing in a
particular direction, a quadrupole corresponds to a double-headed
arrow ($\updownarrow$) not favoring one direction over another, and
therefore it can be aligned, but not oriented.}. In this work we focus on systems possessing cylindrical symmetry, for which only one component $q$ of the quadrupole moment tensor $q_{ij}$ is nonzero.
 The electric field  potential generated by the quadrupole is given by $\phi(\vec{r}) = \frac{q}{4 r^{3}}(3 \cos^{2}\theta -1)$.
 If a second quadrupole with the same alignment $\hat{k}$ is placed at location $\vec{r}$, the resulting interaction energy is $V^{qq}_{cl} = \frac{q}{4} (\hat{k}\cdot \nabla) (\hat{k} \cdot \nabla \phi) = \frac{3q^{2}}{16 r^{5}}(35 \cos^{4}\theta- 30 \cos^{2}\theta +3)$. 
 While the functional form of this potential carries over to the quantum description,  the prefactor of the interaction for two  states has to be obtained via a quantum definition of  the quadrupole moment $\mathbf{q}_2$. The latter is a spherical
tensor of rank two with components defined  in atomic units as $q_{2,M} = - \sum_k
r^2_k C_{2, M} (\theta_k, \phi_k)$, where $(r_k, \theta_k, \phi_k)$
give the coordinates of the $k$-th electron of the particle, and $C_{2,
  M} (\theta_{k}, \phi_{k}) = \sqrt{4\pi/5}\: Y_{2, M} (\theta_{k},
\phi_{k})$ are the reduced spherical
harmonics~\cite{VarshalovichAngMom}. 
 In the angular momentum
basis, $\vert J, M \rangle$, with $M$ being the projection of the
angular momentum, $\mathbf{J}$, onto the quantization axis, the
quadrupole operator couples the states with $\Delta J =0, \pm 2$,
so to first order any state with $J>1/2$ possesses a nonzero
quadrupole moment~\footnote{Note the difference with dipole moments which can be nonzero only for states of indefinite parity (superposition of odd and even $J$'s)}. Thus the value of the quadrupole moment can be
controlled by preparing the particles in a particular $\vert J, M \rangle$-state, or their
combination, using optical or microwave fields. The quadrupole interaction reads
\begin{equation}
\label{Vqq}
	V^{qq} = \frac{\sqrt{70}}{r^5} \sum_{\alpha} (-1)^\alpha C_{4, -\alpha} (\theta, \phi) [\mathbf{q}_2^{(1)} \otimes \mathbf{q}_2^{(2)}]_{4,\alpha},
\end{equation}
where $(r, \theta, \phi)$ gives the vector between
particles, and $[\mathbf{q}_2^{(1)} \otimes
\mathbf{q}_2^{(2)}]_{4,\alpha}$ is a spherical tensor of rank four
formed from two quadrupole moments.  For both
particles prepared in the same $\vert J, M \rangle$ state,
Eq.~(\ref{Vqq}) reduces to $V^{qq} = V (3-30 \cos^2\theta + 35 \cos^4 \theta) / r^5 $, with $V = q^2 3 (J^2 + J -
3M^2)^2/[4(4J^2 + 4J - 3)^2]$, where   $q = 2 \langle 2, 2 \vert q_{2,0} \vert 2, 2 \rangle$, which coincides with the classical definition~\cite{DereviankoPRL01}.
 We note that in the classical limit of $J \to \infty$,  and for $M=J$, the prefactor $V=3q^2/16 $ of the classical expression is recovered. The   interaction can then be attractive or
repulsive depending on the angle $\theta$.

Among the particles for which the quadrupolar moment is known, the 
 most promising candidates for the experimental realization of quadrupolar
quantum gases are metastable alkaline-earth atoms~\cite{JensenPRL11,
  DereviankoPRL01, SantraPRA04, SantraPRA03, BuchachenkoEPJD11} and homonuclear diatomic
molecules~\cite{KreStwFrieColdMolecules, byrdJCP11}. Alkaline-earth atoms, such
as Sr, and some of the rare-earth atoms, such as Yb, can be prepared in metastable $^3P^o_2$ states, whose
lifetime exceeds thousands of seconds~\cite{JensenPRL11,
  DereviankoPRL01, SantraPRA04, SantraPRA03, BuchachenkoEPJD11, NagelPRA03}, by
optical excitation, cf. Fig.~\ref{levels}~(a).  Both bosonic and
fermionic isotopes of Sr and Yb have been brought to quantum
degeneracy~\cite{FukuharPRA07, StellmerPRL09, DeSalvoPRL10,
  FukuharaPRL07}.
  Ultracold homonuclear
molecules, such as Cs$_2$ or Sr$_2$, can be prepared in the absolute
ground state, $^1\Sigma_g^+ (v=0, J=0)$, and then transferred to a
rotational state with $J>0$, using a  Raman transition~\cite{ManaiPRL12, KreStwFrieColdMolecules, byrdJCP11}, cf. Fig.~\ref{levels}~(b). While homonuclear molecules are always bosons, fermionic
quadrupolar molecules can be prepared using distinct isotopes of the same species~\cite{PappPRL06}.  For both
 atoms and molecules the degeneracy of a particular $J$
level can be lifted by an external electric or magnetic field, $\mathbf{F}$. We consider the regime when the quadrupole-quadrupole interactions dominate the behavior of the system, i.e. the electric field $\mathbf{F}$ is too weak to induce a substantial value of a dipole moment, or the particles are prepared in a non-magnetic  Zeeman component. The
typical ``shapes'' of quadrupolar states are exemplified in
Fig.~\ref{levels}~(c). Both atoms and molecules can be prepared in the $\vert 2, 0 \rangle$ ($\vert 2, 2 \rangle$) states using two linearly (circularly) polarized photons; the quadrupole-quadrupole interaction is equal in these cases and is larger than for the $\vert 2, 1 \rangle$ states.

The quadrupole moments for metastable alkaline-earth atoms and
homonuclear molecules are similarly on the order of 10--40
a.u.~\cite{byrdJCP11, JensenPRL11, DereviankoPRL01, SantraPRA04,
  SantraPRA03, BuchachenkoEPJD11}, which gives an interaction
strength, $V^{qq}$, on the order of a few Hz at 266 nm
lattice spacing. Furthermore,  interactions on the order of 1 kHz can be achieved for 100 nm lattice spacings realizable with atoms trapped in nanoplasmonic structures~\cite{GullansArxiv12}. We note that
the dispersion (van der Waals) interaction, $V^\text{dis} \sim r^{-6}$, is $10^2 -
10^3$ times smaller at typical optical lattice
spacings~\cite{SantraPRA04}, therefore the quadrupole-quadrupole
interaction dominates the physics of these systems.  Quantum gases can
be trapped for tens of seconds, so the 
observation of many-body phases generated by these interactions is feasible via the standard techniques, ranging from time-of-flight detection to noise correlation and Bragg spectroscopy.  

\begin{figure}[t]
\includegraphics[width=0.8\linewidth]{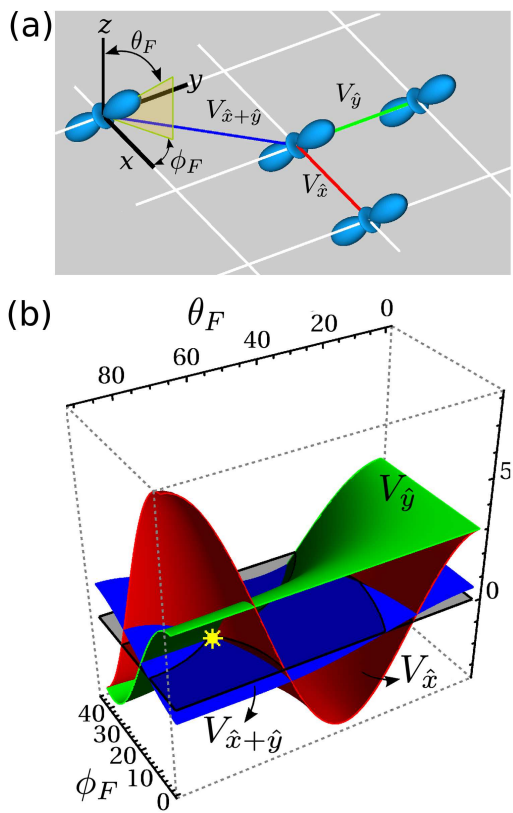}%{Fig22.pdf}
\caption{(a) Schematic representation of quadrupolar fermions on a square
  lattice. Alignment of the quadrupoles is given by the quantization
  axis of the external field ${\bf F}$, pointing along  $\hat{F}=(\theta_F,\phi_F)$. The nearest-neighbor
  interaction is represented by green and red solid lines, while the
  next-nearest neighbor interaction is shown in blue. (b) 3D plot
  showing the interactions $V_{\hat{x}}$ (red), $V_{\hat{y}}$ (green),
  and $V_{\hat{x}+\hat{y}}$ (blue) as a function of the angles
  $(\theta_F,\phi_F)$; ``$*$'' marks the point in the vicinity of
  which both $V_{\hat{x},\hat{y}}$ and $V_{\hat{x}+\hat{y}}$ change the sign.}
\label{pot}
\end{figure}

% edits
\begin{figure*}
\includegraphics[width=0.9\linewidth]{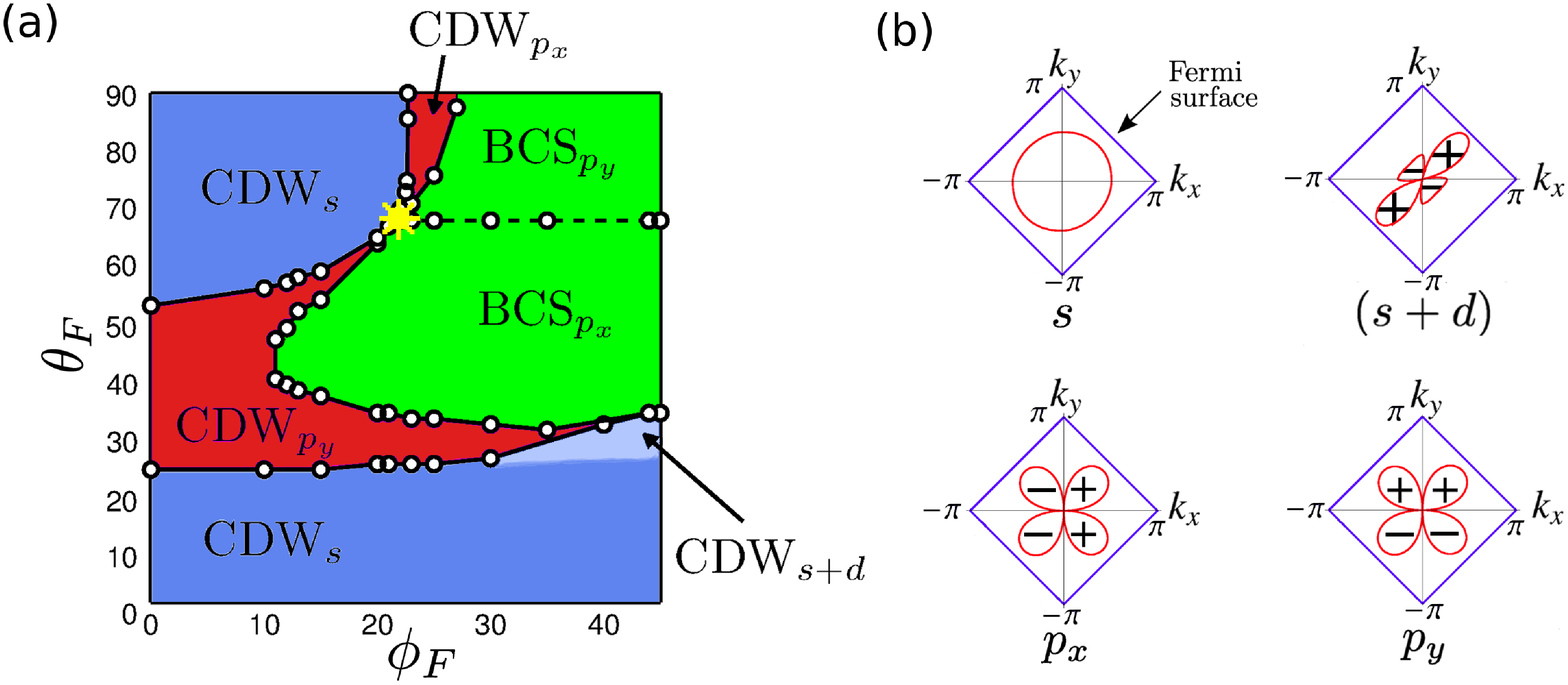}%{Fig33.pdf}
\caption{\label{phasedig} Quantum phase diagram for quadrupolar
  fermions on a square lattice. (a) FRG phase diagram in the weak
  coupling limit, $V/t=0.2$, and at half filling shown as a function
  of the magnetic field direction $\hat{F}=(\theta_F,\phi_F)$.  The
  point marked by ``$*$'', where 5 different phases seem to meet,
  corresponds to $V_{\hat{x}}$, $V_{\hat{y}}$, $V_{\hat{x}+\hat{y}}$
  $\approx 0$ as shown in Fig.~\ref{pot}~(b). This suggests the
    likelihood of atmost three distinct tri-critical points in close
    proximity to each other, which are hard to resolve due to the smallness
    of the couplings. (b) The orbital symmetry of the
  CDW and BCS phases shown in (a) is plotted in red. This symmetry
  corresponds to the symmetry of the most divergent eigenvector of the
  BCS and CDW vertices combined. The $p_x$ ($p_y$) wave CDW has the
  same orbital structure as the $p_x$ ($p_y$) wave BCS.}
\end{figure*}

To illustrate the intriguing many-body effects that can arise in these systems, 
 we investigate the quantum phase diagram of a system of interacting quadrupolar fermions on a  square lattice, at half-filling.  We assume every particle to be 
prepared in state $\vert J, M \rangle$, where $\mathbf{J}$
is the electronic (for atoms) or rotational (for molecules) angular
momentum, and $M$ is the projection of $\mathbf{J}$ on the direction
$\hat{F}=(\theta_F, \phi_F)$ in the laboratory frame given by the
external field $\mathbf{F}$ used to lift the $M$-degeneracy.  
The particles are confined to a lattice  with a lattice constant $a_L$, corresponding to the Hamiltonian:
\begin{equation}
\label{Hamil}
	H=-t \sum_{\langle i,j \rangle} c^\dagger_i c_j + \frac{1}{2} \sum_{i\neq j} V_{ij} c^\dagger_i c_i c^\dagger_j c_j,
\end{equation}
where $t$ represents the nearest-neighbor hopping and $c_i$ is the fermion
annihilation operator at the $i$-th lattice site.
 Throughout the remainder of the paper, we use $a_L$ as a unit of length and   $t$ as a unit of energy.  As schematically shown in Fig.~\ref{pot}~(a), the interaction
strength $V_{ij}$ depends on the orientation of the vector connecting the
quadrupoles, ${\bf r}={\bf r}_i-{\bf r}_j$, relative to the field
direction, $\hat{F}$, via $V_{{\bf r}}\equiv V_{ij}=\langle ij |
V^{qq} | ij \rangle=V[3-30(\hat{r}\cdot \hat{F})^2+35(\hat{r}\cdot
\hat{F})^4]/r^5$. Thus, one can immediately observe that the
interaction between two quadrupoles can be tuned either attractive or
repulsive, by changing the orientation of the external field ${\bf F}$. Fig.~\ref{pot}~(b)
shows the $(\theta_F,\phi_F)$-dependence of the interaction matrix
element between the nearest- and next-nearest neighbors.  The richness
of the quadrupolar interaction becomes apparent in this figure.  There
are several regions in which the signs and the relative magnitudes of
$\{V_{\hat{x}}$, $V_{\hat{y}}$, $V_{\hat{x}+\hat{y}}\}$ show
distinctive characteristics. For example, in the region
$(\theta_F\lesssim 25^\circ,0^\circ\le \phi_F\le 45^\circ)$, both
nearest- and next-nearest neighbor interactions are repulsive, while
they all become attractive in the region
$(30^\circ\lesssim\theta_F\lesssim 60^\circ, \phi_F\sim
45^\circ)$. Furthermore, one can identify finite regions where either
one or two of $\{V_{\hat{x}}$, $V_{\hat{y}}$, $V_{\hat{x}+\hat{y}}\}$
is attractive while the rest is repulsive.

Interactions of opposite sign can result in competition between 
 quantum phases of different symmetry,  resulting in frustration.
%Even with the same sign, they may still compete and cause 
%for example frustration.
 Thus, fermions with dominant quadrupolar interactions 
provide an interesting  setup for studying many-body physics 
with competing phases. For example, in the vicinity of
$(90^\circ,45^\circ)$ both
$V_{\hat{x}}$ and $V_{\hat{y}}$ are
attractive, while $V_{\hat{x}+\hat{y}}$ is repulsive (see Fig.~\ref{pot}). On general grounds, one would expect
 a BCS type ground state
resulting from condensation of Cooper pairs
due to the attractive $V_{\hat{x}}$ and $V_{\hat{y}}$ couplings. However, the repulsive
$V_{\hat{x}+\hat{y}}$ interaction, if significant, may lead to the insurgence of some other phase, and therefore needs to be quantitatively accounted for.
%needs to be quantitatively accounted for. 
As another intriguing example, in the vicinity of $(40^\circ, 5^\circ)$,
$V_{\hat{x}}$ is strongly attractive while $V_{\hat{y}}$ 
is strongly repulsive. As we show below, the ground state in this region 
is neither a BCS state nor conventional charge density wave (CDW). 
These two examples show that 
the actual ground state may be of an unexpected nature. 
Exposing the true ground state thus demands a
theory that  is (i) unbiased with respect to the initial ansatz, and (ii)
includes fluctuations.

Issue (ii) can be adequately addressed within the
renormalization group (RG) analysis at weak couplings, where the low energy physics near
the Fermi surface is extracted by successively integrating out the high
energy degrees of freedom \cite{ShankarRMP94}. In order to satisfy
criterion (i), we employ the exact (or ``functional'') renormalization group (FRG) which keeps
track of all the interaction vertices, including both the particle-particle
and particle-hole channels, and treats all instabilities 
on equal footing  \cite{ZanchiPRB00}\footnote{Details on the FRG approach are provided in the supplemental online material.}.

The FRG phase diagram, Fig.~\ref{phasedig}~(a), features several
different kinds of BCS and CDW phases with symmetry indicated by the the polar plots of Fig.~\ref{phasedig} (b).  CDW$_s$ is  a CDW phase with a checkerboard modulation of on-site
densities, occurring in regions where the
repulsive interaction between nearest neighbors dominates, see Fig.~\ref{pot}. This
happens for all values of $\phi_F$ when $\theta_F\lesssim 25^\circ$,
and also for $\phi_F\lesssim 22^\circ$ at large $\theta_F\gtrsim
60^\circ$. In addition, two novel types,
CDW$_{p_{x}}$ and CDW$_{p_{y}}$, are present.  
They correspond to a checkerboard modulation of the effective hopping 
between nearest neighbors along the $x$ and $y$ direction respectively, 
i.e., $\langle c^\dagger_i c_{j}\rangle$ with ${\bf r}_{i}-{\bf r}_j=\hat{x}\,\, \text{or}\,\, \hat{y}$,  
with the average taken over the many-body ground state. 
We refer to these phases as to bond order solids (BOS).
In comparison, 
the  $s$-wave CDW order corresponds to modulations of $\langle c^+_i c_i\rangle$.
Furthermore, we find a small region  of CDW$_{s+d}$ that involves a mixture 
of extended $s$- and $d$-waves. 
Together they give rise to a checkerboard modulation of effective hopping 
between the next-nearest neighbor sites. The 
CDW$_{p_{x}}$,  CDW$_{p_{y}}$, and CDW$_{s+d}$ can be thought of as a 2D
generalization of the bond-order-wave phase occurring in the extended Hubbard
model in one dimension
\cite{NakamuraPRB00, SenguptaPRB05, TamPRL06}. While  BOS is expected for dipolar fermions in 2D \cite{BhongalePRL12}, it occupies a  significantly larger region of the
parameter space for quadrupolar interactions (e.g., it is stabilized
as soon as $\theta_F$ approaches $25^\circ$).  Moreover, the angular
dependence of quadrupolar interactions is substantially more complex,
resulting in two BOS phases of $p_y$ and $p_x$ symmetries, appearing
at small and large $\theta_F$, respectively. Interestingly, these two
phases occur in the regions where $V_{\hat{x}}$ and $V_{\hat{y}}$ are
comparable in magnitude but opposite in sign, i.e.,
CDW$_{p_{x}(p_{y})}$ is stabilized when $V_{\hat{x}(\hat{y})}$ is
repulsive while $V_{\hat{y}(\hat{x})}$ is attractive, cf.\
Fig.~\ref{pot}~(b).  Thus, quadrupolar Fermi gases are well suited
for exploring the properties of nonzero angular momentum (i.e.\
unconventional) density wave phases.

Finally there are two BCS phases, which mostly occur where both $V_{\hat{x}}$ and
$V_{\hat{y}}$ are attractive. Our FRG analysis shows that the BCS
phase can be stable even though the next-nearest neighbor interaction
is weakly repulsive.  We find that the symmetry of the BCS order
parameter is $p_x$ or $p_y$, depending on whether $V_{\hat{x}}$ or $V_{\hat{y}}$ is more
attractive.  Along the line of $\theta_F\sim 65^\circ$ these two BCS
phases are degenerate.  This raises the possibility of realizing
$p_x+ip_y$ topological superfluid order. By analogy
with the proposal of Ref.~\cite{CooperPRL09}, using an AC field
to periodically modulate the direction of $(\theta_F,\phi_F)$, one can
lift the degeneracy and engineer the chiral $p_x+ip_y$ state.

In conclusion, we have shown that ultracold Fermi gases with quadrupole-quadrupole interactions can be used to study unconventional BCS, CDW, and topological phases, and gain insight into the physics of competing ground states. While we have
 focused on the specific case of a square lattice at half-filling, the functional RG
methods of this work can be applied to study
other fillings and lattice geometries. Temperatures
achieved for degenerate Fermi gases of alkaline-earth atoms in
experiment are $T=0.26~T_F$ and $T=0.37~T_F$ respectively~\cite{DeSalvoPRL10,
  FukuharaPRL07}. The optimal $T_c$ for the CDW and BCS
phases predicted here is estimated to be on the order of $0.03~T_F$,
for intermediate couplings, $V\sim t$. Thus these many-body phases
seem to be within experimental reach in the near future.

 Since
quadrupolar interactions occur in numerous subfields of physics, from
molecular photofragmentation~\cite{AlexanderPCCP05} and structure of
$f$-electron compounds~\cite{ItoPRB11} to nuclear
reactions~\cite{BlattWeisskopf79} and gravitation of black
holes~\cite{BambiPRD11}, the proposed quantum simulation platform can in principle be applied beyond the many-body physics
of fermionic gases.
Finally, we  note that ground-state  atoms can be provided with
significant quadrupole moments by means of Rydberg dressing, i.e.
admixing a highly-excited electronic state possessing a large
quadrupole moment with far-detuned laser light~\cite{FlanneryJPB05,
SaffmanMolmerRMP10}.

%\begin{acknowledgement}
We are grateful to Charles Clark, Robin Cot\'e, Hendrik Weimer, and Shan-Wen Tsai for 
discussions. S.B. and E.Z. are supported by NSF (PHY-1205504) and NIST
(60NANB12D244). L.M. acknowledges support from the
Landesexzellenzinitiative Hamburg, which is financed by the Science
and Research Foundation Hamburg and supported by the Joachim Herz
Stiftung, and from the Deutsche Forschungsgemeinschaft under SFB 925.  
M.L. acknowledges support from NSF through a grant for the
Institute for Theoretical Atomic, Molecular, and Optical Physics at
Harvard University and Smithsonian Astrophysical Observatory.
S. F. Y. acknowledges financial support from the National Science Foundation and Air Force
Office of Scientific Research. 
%\end{acknowledgement}

\newpage

\section{Supplemental Material} 

%\begin{abstract}
%  Here we outline the functional renormalization group (FRG) method
%  applied to correlated fermions.
%\end{abstract}

\subsection{Functional Renormalization Group}

Renormalization group (RG) is a symmetry transformation allowing
to map the Hamiltonians or actions defined in a certain phase
space to those in the same phase space. If one represents the
initial action as a point in the coupling constant space, this point
will move under the RG transformation to another point in the same
space, implying a {\it flowing} coupling constant as the RG
transformation is performed repeatedly. This automatically gives rise
to the notion of a fixed point where the flow stops implying constancy
of couplings under the RG transformation. Furthermore, the RG transformation, associated with a finite cutoff,
allows for separation between the modes of interest (low energy
degrees of freedom) and the rest, by decreasing the cutoff
accompanied by a suitable change in couplings such that the theory is
invariant.

The ground state of a non-interacting Fermi system is a filled Fermi
sea bounded by the Fermi surface (FS). Such a non-interacting ground
state has gapless excitations corresponding to promoting a fermion
from just below the FS to above the FS. However, if some perturbation is
added to the free system, will the system develop a gap or remain
gapless? This connection, $T=0$ {\it gapless} Fermi liquid as a RG fixed
point, and the various {\it gapped} quantum phases as Fermi surface
instabilities associated with diverging RG flow, was originally
introduced by Shankar \cite{ShankarRMP94}. The method used here, the
functional renormalization group (FRG), is a generalization of
Shankar's renormalization-group to an arbitrary Fermi surface
\cite{ShankarRMP94}. The Kadanoff-Wilson mode elimination (developed by
Shankar for 2D fermions) applied to the effective action with only two
particle interaction retains only strictly logarithmic contributions
to the flow. Therefore, {\it e.g.}, even if the nesting is good but
not perfect, the corresponding singularity in the the $p-h$ channel is
destroyed reducing the RG-flow to zero. As one can imagine, this could
be a serious issue when analyzing situations where the low energy
physics is an outcome of the interplay between several competing
channels. As a remedy, Zanchi and Schulz derived the functional
renormalization group technique \cite{ZanchiPRB00} as a generalization of
the Kadanoff-Wilson-Polchinski group \cite{PolchinskiNuclPhysB84} that was originally
formulated in the context of quantum fields with one single zero
energy point in momentum space.

\begin{figure}[b]
  \includegraphics[scale=.5]{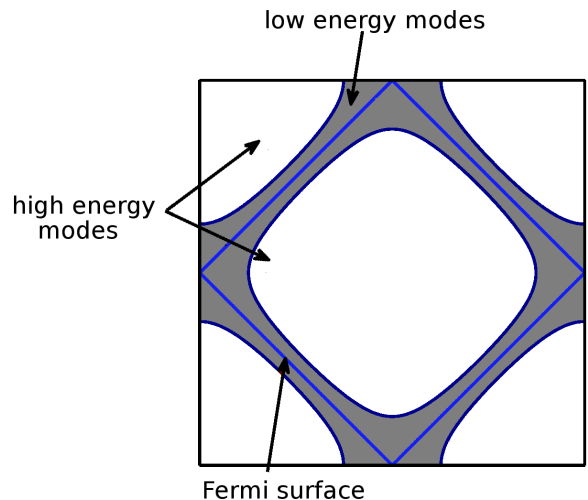}
  \caption{Schematic indicating the low energy modes in the vicinity
    of the Fermi surface for 2D fermions on square lattice.}
\label{fermisurf}
\end{figure}

The starting point of the renormalization group for a many-fermion
model Hamiltonian is the partition function given by
\begin{eqnarray}
\mathcal{Z}=\int \sum_{\gamma,\sigma}\mathcal{D}\bar{\psi}_{\gamma,\sigma}\mathcal{D}\psi_{\gamma,\sigma} e^{S[\bar{\psi}_{\gamma,\sigma},\psi_{\gamma,\sigma}]} \nonumber
\end{eqnarray}
where the integration is carried out over Grassmann variables
$\{\bar{\psi}_{\gamma,\sigma},\psi_{\gamma,\sigma}\}$ for all fermions
in the Brillouin zone where the index $\gamma$ represents
energy-momentum vector $(\omega_n,{\bf k})$ and $\sigma$ is the spin.
Since we are interested in the low energy physics, the goal is to find
an effective action that depends only on the low energy degrees of
freedom close to the Fermi surface. If we assume that the slow modes,
as illustrated in supplementary figure \ref{fermisurf}, are in a shell
$\pm\Lambda$ about the Fermi surface, then the action $S$ maybe
decomposed as
\begin{eqnarray}
S\{\psi\}&=&S_0\{\psi_{<}\}+S_0\{\psi_{>}\}+S_I\{\psi_{<},\psi_{>}\},\nonumber
\end{eqnarray}
where $S_0$ is the non-interacting piece and $S_I$ comes from the
two-particle interaction. Now, from the definition of the partition
function, it is easy to see that the low energy effective action can
be formally written as a partial trace over the high energy degrees of
freedom as:
\begin{eqnarray}
S_{\Lambda}\{\psi_{<}\}&=&\ln \int\mathcal{D}\bar{\psi}_{>}\mathcal{D}\psi_{>}e^{S\{\psi_{<},\psi_>\}}\nonumber
\end{eqnarray}
Further, this effective action can always be decomposed into three
terms,
\begin{eqnarray}
S_{\Lambda}\{\psi_{<}\}&=&S\{\psi_{<}\}+\Omega_{>}+\delta S\{\psi_{<}\}\nonumber
\end{eqnarray}
Here $\Omega_{>}$ is the grand potential (times $\beta$) of the fast
fermions as if they were decoupled from the slow ones, and $\delta
S_{<}$ is the correction arising from coupling to the fast ones. This
correction essentially gives rise to terms that are quadratic,
quartic, etc, in field variables, thus renormalizing the corresponding
self energy, two-point vertex, etc, respectively. Now all that remains
is to rescale the momenta and field variables so that that
$S_{\Lambda}$ describes the low energy physics over the same kinematic
range as before, allowing the proper comparison of the old and new
couplings.

% \begin{figure}[t]
%   \includegraphics[scale=.4]{feyn_frg.eps}
%   \caption{Possible vertex diagrams with one internal line (shown in
%     blue), that is integrated over the shell $\Lambda$. a)
%     $\Gamma_{2n}$ for $n=2$, b) and c) correspond to possible vertex
%     diagrams for $n>2$.}
% \label{feyn}
% \end{figure}

However, the problem that remains to be addressed is to decide at what
order the effective action can be truncated so that the relevant low
energy physics is captured. For fermions near the Fermi surface, even
if the bare couplings are small, some of the diagrams, for example the
particle-hole bubble, can acquire large values depending on the shape
of the Fermi surface. Thus, in principle all $p-h$ diagrams will have
to be summed over. Such a sum (one example is the random phase approximation (RPA)) can only be
controllably carried out in a limited number of situations and even
those are rendered inefficient if the couplings are large.

The FRG provides a tractable way of book keeping the order of diagrams
by defining an infinitesimal group transformation
\begin{equation} 
S \equiv S_{\Lambda_0}\rightarrow S'_{\Lambda_0 e^{-\ell}} \nonumber
\end{equation}
where $\Lambda_0$ is the initial cutoff and $\Lambda=\Lambda_0
e^{-\ell}$. Thus at each step $\Lambda d\ell$ modes are eliminated at
a shell size $\Lambda$. In terms of diagrams this procedure allows for
classification of diagrams in order of $d\ell$. However, since every
internal line is constrained to a shell, a diagram of order
$(d\ell)^m$ will have $m$ internal lines. Now, since $d\Lambda$ is
infinitesimal, only diagrams with one internal line dominates at every
step of RG. This allows for grouping diagrams with equal number of
external legs, resulting in the following Polchinski equation (flow
equation) for the vertex $\Gamma_{2n}^{\ell}$
\begin{widetext}
\begin{eqnarray}
\frac{\partial}{\partial \Lambda_{\ell}}\Gamma_{2n}^{(\ell)}(K_1,K_2,..,K_{2n})&=&\sum_{I_1,I_2}T\sum_{\omega_n}\int_{d\Lambda}d^2k\Gamma_{2p}^{(\ell)}(-K,I_1)G_{\ell}(K)\Gamma_{2q}^{(\ell)}(K,I_2)\nonumber\\
&&-T\sum_{\omega_n}\int_{d\Lambda}d^2k\Gamma_{2(n+1)}^{(\ell)}(K,K_1,..K_n,K,K_{n+1},..,K_{2n})G_{\ell}(K),\nonumber
\end{eqnarray}
\end{widetext}
where $I_1\cup I_2=\{K_1,K_2,...,K_{2n}\}$ and the propagator is defined at each step by 
\begin{equation}
G_{\ell}(K)=\left[\Gamma_2^{(\ell)}(K)\right]^{-1}\nonumber
\end{equation}
The possible vertex
diagrams with $2n$ external lines and one internal line are shown in
Supplementary Figure ~\ref{feyn}.
\begin{figure}[t]
  \includegraphics[scale=.5]{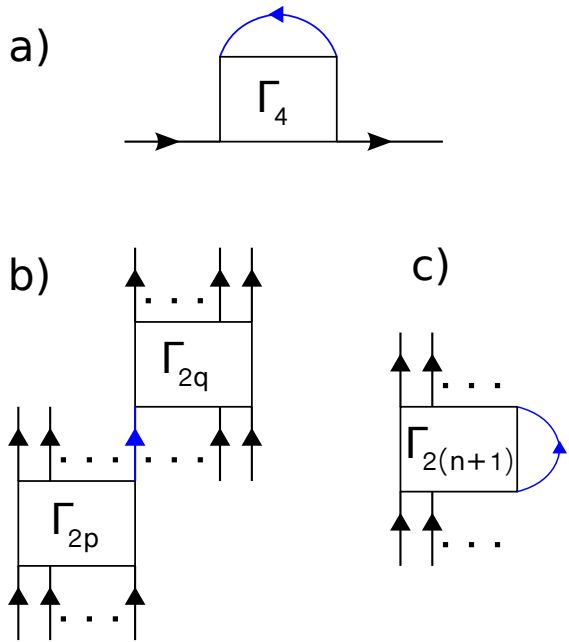}
  \caption{Possible vertex diagrams with one internal line (shown in
    blue), that is integrated over the shell $\Lambda$. a)
    $\Gamma_{2n}$ for $n=2$, b) and c) correspond to possible vertex
    diagrams for $n>2$.}
\label{feyn}
\end{figure}

\subsection{Numerical Implementation}
The Fermi surface (FS) of the non-interacting system of quadrupoles on
a square lattice is a nested square at half filling.  We discretize
the FS into $N=32$ patches with momenta ${\bf k}_i$,
$i\in\{1,2,..N\}$. The 4-point interaction vertex as a function of
three momenta $\Gamma^{(\ell)}_4({\bf k}_1,{\bf k}_2,{\bf k}_3)\equiv
U_{\ell}({\bf k}_1,{\bf k}_2,{\bf k}_3)$ is represented as a
$N^3\times N^3$ matrix. We start with the bare vertices defined on the
discretized Fermi surface, $U_{\ell=0}({\bf k}_i,{\bf k}_j,{\bf
  k}_l)=V_{{\bf k}_i-{\bf k}_l}$, where $V_{{\bf q}}$ is the
quadrupole interaction in momentum space, i.e., the lattice Fourier
transform of $V_{{\bf r}}$ given in the main text, with proper
anti-symmetrization required by Fermi statistics.  The momentum
dependence of the interaction is fully taken into account.  At each
FRG step $\ell$, the renormalized vertices $U_{\ell}({\bf k}_i,{\bf
  k}_j,{\bf k}_l)$ are calculated at one-loop level, truncating the
effective action at the six fermion terms and neglecting self-energy
correction \cite{ZanchiPRB00}. The renormalized vertices for particular
channels, e.g. $U_{\ell}^{\text{CDW}}({\bf k}_1,{\bf k}_2)\equiv
U_{\ell}({\bf k}_1,{\bf k}_2,{\bf k}_1+{\bf Q})$ and
$U_{\ell}^{\text{BCS}}({\bf k}_1,{\bf k}_2)\equiv U_{\ell}({\bf
  k}_1,-{\bf k}_1,{\bf k}_2)$, are then extracted by appropriately
choosing the in-coming and out-going momenta and ${\bf Q}=(\pm \pi,\pm \pi)$ is the nesting vector. The channel with the
most divergent eigenvalue represents the dominant instability of the
FS, thereby indicating the nature and symmetry of incipient order
parameter.

% Within our implementation of FRG, the shape of the Fermi surface is
% fixed, certainly a good assumption at weak couplings. Also, other
% higher order effects that can be safely neglected in the weak coupling
% limit.  Similar arguments were made in many FRG studies e.g. on the
% Hubbard model \cite{zanchi}. The FRG phase diagram is most reliable
% and trustworthy in the weak coupling regime. The phase diagram is
% expected to be modified at stronger interactions.

 \bibliography{References_library}

\end{document}